\documentstyle[prl,twocolumn,aps]{revtex}
\begin{document}

\title{Exact Single-Particle Green Functions of Fermi Systems Without
 Using Bosonization or Are Hartree-Fock and Random Phase Approximations
 'Controlled Approximations' ?}

\author{Girish S. Setlur and Yia-Chung Chang}
\address{Department of Physics and Materials Research Laboratory,\\
 University of Illinois at Urbana-Champaign , Urbana Il 61801}
\maketitle

\begin{abstract}
 In this article, we revisit the question of the validity of Hartree-Fock and
 random-phase approximations. We show that there is a
 connection between the two and while
 the RPA as it is known in much of the physics literature is of limited 
 validity, there is a generalised sense in which the random phase approximation
 is of much wider applicability including to systems that do not possess
 Fermi surfaces. The main conclusion is that the Hartree-Fock approximation
 is a mean-field idea applied to the density operator, and the
 random-phase approximation is a mean-field idea 
 applied to the number operator. The generalised RPA is used to compute 
 single-particle properties such as momentum distribution and spectral
 functions. It is found that we have to go beyond the generalised RPA
 and include fluctuations in the momentum distribution in order to 
 recover a nonzero imaginary part of the one-particle self energy, which
 is also explicitly computed, all this in any number of spatial dimensions
 and no bosonization is needed. 
\end{abstract}

\hspace{13in}

 Both the Hartree-Fock\cite{Hartree}
 and the Random Phase approximations(RPA)\cite{Bohm}
 are widely used in the physics literature to study Fermi systems.
 The Bogoliubov theory\cite{Bogoliubov}
 is the analog of the random-phase approximation for Bose systems.
 This latter fact has been
 demonstrated in detail in our article\cite{Setlur}.
 The analysis presented here could be repeated for Bose systems as well.
 Here we try and address the
 question of validity of the Hartree-Fock approximation and the RPA.
 We show that the mean-field approximation carried out on the density operator
 is the Hartree-Fock approximation.
 The RPA manifests itself as mean-field theory applied not
 to the density operator but to the number operator. The density operator
 measures how the electrons are distributed in real space where as the number
 operator measures how the electrons are distributed in momentum space. 
 Just as the Hartree-Fock approximation
 is valid when fluctuations in the density
 of electrons at each point in real space is small, the RPA is valid when
 fluctuations in the momentum distribution of electrons
 is small compared with the average momentum distribution
 which measures the probability of an electron to possess 
 a given momentum.
  
 Let us try and write down some representative examples when RPA and 
 Hartree-Fock approximations are used.
  Let us take for example, the jellium model
 \cite{Mahan}. 
\begin{equation}
H = \sum_{ {\bf{k}} }\epsilon_{ {\bf{k}} }c^{\dagger}_{ {\bf{k}} }c_{ {\bf{k}} }
+ \sum_{ {\bf{q}} \neq 0 }
\frac{v_{ {\bf{q}} } }{2V}(\rho_{ {\bf{q}} }\rho_{ -{\bf{q}} } - N)
\end{equation}
 If we apply the mean-field idea on the density, that is, replace
 $ \rho_{ {\bf{q}} } $ by $ \langle \rho_{ {\bf{q}} } \rangle $
 then we get a hamiltonian that does not involve any coulomb interaction
 at all(apart from an additive constant).
\[
H = \sum_{ {\bf{k}} }\epsilon_{ {\bf{k}} }c^{\dagger}_{ {\bf{k}} }c_{ {\bf{k}} }
+ \sum_{ {\bf{q}} \neq 0 }
\frac{v_{ {\bf{q}} } }{2V}(\langle \rho_{ {\bf{q}} } \rangle\rho_{ -{\bf{q}} } - N)
\]
\begin{equation}
= \sum_{ {\bf{k}} }\epsilon_{ {\bf{k}} }c^{\dagger}_{ {\bf{k}} }c_{ {\bf{k}} }
-N \sum_{ {\bf{q}} \neq 0 }
\frac{v_{ {\bf{q}} } }{2V} 
\end{equation}
 Therefore this approximation is bad in the extreme. However the
 random-phase approximation is still valid for this system.
 It is the mean-field idea
  applied not to the density but to the number operator.
 For this we have to rewrite the full hamiltonian given below,
\begin{equation}
H = \sum_{ {\bf{k}} }\epsilon_{ {\bf{k}} }c^{\dagger}_{ {\bf{k}} }c_{ {\bf{k}} }
+ \sum_{ {\bf{q}} \neq 0 }
\frac{ v_{ {\bf{q}} } }{2V}\sum_{ {\bf{k}} , {\bf{k}}^{'} }
c^{\dagger}_{ {\bf{k}}+{\bf{q}}/2 }c^{\dagger}_{ {\bf{k}}^{'}-{\bf{q}}/2 }
c_{ {\bf{k}}^{'}+{\bf{q}}/2 }c_{ {\bf{k}}-{\bf{q}}/2 }
\end{equation}
This has to be replaced by,
\begin{equation}
H = 
H_{0}
- \sum_{ {\bf{q}} \neq 0 }
\frac{v_{ {\bf{q}} } }{2V}\sum_{ {\bf{k}} \neq  {\bf{k}}^{'} }
c^{\dagger}_{ {\bf{k}}+{\bf{q}}/2 }
c_{ {\bf{k}}^{'}+{\bf{q}}/2 }
 c^{\dagger}_{ {\bf{k}}^{'}-{\bf{q}}/2 }c_{ {\bf{k}}-{\bf{q}}/2 } 
\end{equation}
where
\begin{equation}
H_{0} = \sum_{ {\bf{k}} }
\epsilon_{ {\bf{k}} }c^{\dagger}_{ {\bf{k}} }c_{ {\bf{k}} }
- \sum_{ {\bf{q}} \neq 0 }
\frac{ v_{ {\bf{q}} } }{2V}\sum_{ {\bf{k}} }
n_{ {\bf{k}}+{\bf{q}}/2 } n_{ {\bf{k}}-{\bf{q}}/2 }
\end{equation}
 where $ n_{ {\bf{k}} } = c^{\dagger}_{ {\bf{k}} }c_{ {\bf{k}} } $.
Let us write,
\begin{equation}
 n_{ {\bf{k}} } = \langle n_{ {\bf{k}} } \rangle
  + \delta  n_{ {\bf{k}} } 
\end{equation}
 We plug the above decomposition into $ H_{0} $ and find that 
 if we neglect terms quadratic in the fluctuations, we get what
 we are after, namely the generalised RPA called for simplicity as just RPA.
\begin{equation}
H_{RPA} = E^{'}_{0} + \sum_{ {\bf{k}} }{\tilde{ \epsilon }}_{ {\bf{k}} }
c^{\dagger}_{ {\bf{k}} }c_{ {\bf{k}} }
\end{equation}
\begin{equation}
{\tilde{ \epsilon }}_{ {\bf{k}} } = \epsilon_{ {\bf{k}} }
 - \sum_{ {\bf{q}} }\frac{ v_{ {\bf{q}} } }{V} \langle n_{ {\bf{k-q}} } \rangle
\end{equation}
The average occupation is
\begin{equation}
\langle n_{ {\bf{k}} } \rangle = \frac{1}
{exp(\beta({\tilde{ \epsilon }}_{ {\bf{k}} } -\mu)) + 1}
\label{SELF1}
\end{equation}
The chemical potential $ \mu $ has to be fixed by making sure that,
\begin{equation}
 \sum_{ {\bf{k}} } \langle n_{ {\bf{k}} } \rangle  = \langle N \rangle
\end{equation}
 At zero temperature $ \mu = \epsilon_{F} $,
 this quantity is equal to the usual Fermi energy
 when $ v_{ {\bf{q}} } = 0 $.
\begin{equation}
  \langle n_{ {\bf{k}} } \rangle = \theta(\epsilon_{F}
-{\tilde{ \epsilon }}_{ {\bf{k}} }) =  \theta(\epsilon_{F}
- \epsilon_{ {\bf{k}} }+ \sum_{ {\bf{q}} \neq 0 }
\frac{ v_{ {\bf{q}} } }{V}\langle n_{ {\bf{k}}-{\bf{q}} } \rangle)
\label{SELF2}
\end{equation}
 and $ \theta $ is the Heaviside step function.
 We can now demonstrate that the
 generalised RPA dielectric function(the Lindhard dielectric function
 \cite{Lindhard} being a weak coupling limit of this)
 may be recovered using the following procedure.
 If one considers an extremely weak external perturbation applied to
 the system and
 follows the discussion in Mahan \cite{Mahan}
 one arrives at the following formula for the dielectric function,
\begin{equation}
\epsilon_{RPA}({\bf{q}},\omega) = 
 1 + \frac{ v_{ {\bf{q}} } }{V}\sum_{ {\bf{k}} }
\frac{ \langle n_{ {\bf{k+q/2}} } \rangle 
- \langle  n_{ {\bf{k-q/2}} } \rangle }
{ \omega - {\tilde{ \epsilon }}_{ {\bf{k+q/2}} } 
+ {\tilde{ \epsilon }}_{ {\bf{k-q/2}} } }
\label{GENRPA}
\end{equation}
 The only point to bear in mind is that we have to use the full interacting
 momentum distribution rather than just the noninteracting value.
 The above formula differs from the traditional RPA in two respects. 
 First, we have the full interacting momentum distribution in the numerator.
 This has to be determined self-consistently by an equation such as
 Eq.(~\ref{SELF1}) or  Eq.(~\ref{SELF2}).
 While this feature may be found in our earlier work\cite{Setlur},
 the second feature is new. The denominator
 contains the renormalised dispersion $ {\tilde{ \epsilon }}_{ {\bf{k}} } $
 rather than the parabolic $ \epsilon_{ {\bf{k}} } $. Thus we can see
 that there is a whole new set of approximations that go beyond the RPA.
 While none of these revelations may come as a surprise to the reader,
 it should serve as a reminder that even our most cherished approximations
 may not be controlled in any sense of the term.
 It is more likely that
 they were the first to appear in the literature 
 and probably the easiest ones to use thus explaining their popularity. 
 The dielectric function written down above has the attractive feature of
 reducing to the familiar Lindhard dielectric function\cite{Lindhard}
 for extremely weak
 coupling and at the same time giving us something very different for
 stronger coupling. The only drawback of the above approach is that if we
 compute the one-particle Green functions we find that the imaginary
 part vanishes identically. This is unfortunate and we have to do better
 in order capture lifetime effects. The RPA hamiltonian neglects 
 fluctuations in the momentum distribution of the electrons. In order
 to recover a finite lifetime of single particle excitations,
 we find that it is important to  study the generalised $ H_{0} $
 rather than the more simple $ H_{RPA} $.
 The fluctuations in the momentum
 distribution may be related to the mean by the following observation.
 Define,
 $ N({\bf{k}}, {\bf{k}}^{'}) = \langle n_{ {\bf{k}} }n_{ {\bf{k}}^{'} } \rangle - \langle n_{ {\bf{k}} } \rangle  \langle n_{ {\bf{k}}^{'} } \rangle $
. The fluctuation in the number operator is
 $ N({\bf{k}}, {\bf{k}}) =  \langle n^{2}_{ {\bf{k}} } \rangle - \langle n_{ {\bf{k}} } \rangle^{2} $ .
 Since $ n^{2}_{ {\bf{k}} } = n_{ {\bf{k}} }  $ for fermions, we have
\begin{equation} 
 N({\bf{k}}, {\bf{k}})
 = \langle n_{ {\bf{k}} } \rangle(1-\langle n_{ {\bf{k}} } \rangle) 
\label{INTEGRALEQ}
\end{equation}  
 Therefore, we may conclude that any nonideal momentum distribution fluctuates
 (there are however, pathological exceptions, see footnote\cite{footnote}).
 In fact, a very nonideal momentum distribution such as one for which 
 $ \langle n_{ {\bf{k}} }  \rangle = 0.5 $ for most momenta has
 the largest fluctuation. When dealing with nonideal
 systems, we are obliged to consider fluctuations in the momentum
 distribution. In order to study lifetime effects therefore, we have to include
 the full $ H_{0} $. This may be written more transparently as
 (apart from additive constants) $ H _{0} = H_{RPA} + H_{fl} $,
\begin{equation}
H_{fl} = -\sum_{ {\bf{q}} \neq 0 }\frac{ v_{ {\bf{q}} } }{2V}
\sum_{ {\bf{k}} } \delta n_{ {\bf{k}} + {\bf{q}}/2 } 
\delta n_{ {\bf{k}} - {\bf{q}}/2 }
\end{equation}
 The full Fermi propagator may be evaluated
 by treating the fluctuation part as a perturbation and using the
 functional methods of Schwinger illustrated brilliantly by Kadanoff and
 Baym\cite{Baym}. 
 Before we plunge into the details it is important to keep in mind
 that the mean also changes when we consider fluctuations. That is,
\begin{equation}
\langle n_{ {\bf{k}} } \rangle = \langle n_{ {\bf{k}} } \rangle_{RPA}
 + \langle n_{ {\bf{k}} } \rangle_{fl}
\end{equation}
 here $ \langle n_{ {\bf{k}} } \rangle_{RPA} $ is given by
 Eq.(~\ref{SELF1}) or Eq.(~\ref{SELF2}) the rest is nonzero only
 when the fluctuation in the momentum distribution is large
 (which is, unfortunately, almost always the case when
 $ \langle n_{ {\bf{k}} } \rangle_{RPA} $ is nonideal)
 Therefore, now, $ \delta n_{ {\bf{k}} } $ refers to fluctuation around
 the full average rather than the RPA average.
 The final answers are given below, and it is hoped that the reader can
 rederive them using the references quoted in the bibliography
  (mainly \cite{Mahan}, \cite{Baym}).
 We shall adhere to the notation of Kadanoff and Baym \cite{Baym}.
 In their notation the final answers for the single-particle
 Green functions are as follows(we assume in the following that
 $ F({\bf{p}}) \neq 0 $, however one may investigate the limit
 $  F({\bf{p}}) \rightarrow 0 $, here $ z_{n} = (2n+1)\pi/\beta $),
\begin{equation}
G_{n}({\bf{k}}) = \frac{1}{iz_{n} - {\tilde{\epsilon}}_{ {\bf{k}} } + \mu
 - \Sigma_{n}({\bf{k}}) }
\end{equation}
and 
\begin{equation}
\Sigma_{n}({\bf{k}}) = G_{n}({\bf{k}})F({\bf{k}})
\label{SELF}
\end{equation} 
\begin{equation}
F({\bf{k}}) = \sum_{ {\bf{q}},{\bf{q}}^{'} \neq 0 }
\frac{ v_{ {\bf{q}} }v_{ {\bf{q}}^{'} } }{V^{2}}
N({\bf{k-q}},{\bf{k}}-{\bf{q}}^{'})
\end{equation}
 From Eq.(~\ref{SELF}) we may obtain the real and imaginary parts of the
 retarded self-energy, and from there the spectral function and the 
 collision rates\cite{Baym}.
\begin{equation}
\Gamma({\bf{p}},\omega) = \sqrt{-\kappa({\bf{p}},\omega)}
\end{equation}
 Similarly,
\begin{equation}
A({\bf{p}},\omega) = \frac{ \sqrt{-\kappa({\bf{p}},\omega)} }{F({\bf{p}})}
\end{equation}
 if $ \kappa({\bf{p}},\omega) = (\omega-{\tilde{\epsilon}}_{ {\bf{p}} }+\mu)^{2}-4F({\bf{p}}) < 0 $
 and both are zero otherwise(when $ F({\bf{p}}) \neq 0 $).
 It can be seen that the spectral function is peaked around 
 $ {\tilde{\epsilon}}_{ {\bf{p}} } - \mu $ with a width of the order
 of $ 2\sqrt{ F({\bf{p}}) } $, and the collision
 rate is vanishingly small for those values of $ {\bf{p}} $
 for which $ F({\bf{p}}) $ is close to zero.
 It may be shown that the momentum
 distribution including possible fluctuations is given by,
\begin{equation}
\langle n_{ {\bf{p}} } \rangle = (\frac{2}{\pi})
\int^{\pi/2}_{-\pi/2}d\theta\mbox{       }cos^{2}\theta\mbox{         }
\frac{1}{e^{\beta({\tilde{\epsilon}}_{ {\bf{p}} }-\mu)}
e^{2\beta \sqrt{ F({\bf{p}}) }sin \theta}
 + 1}
\label{FLUCMOM}
\end{equation}
 Again, it may be seen quite easily that Eq.(~\ref{FLUCMOM}) is identical to
 Eq.(~\ref{SELF1}) when fluctuations in the momentum distribution
 are ignored.
 The only sticking point now is the computation of the fluctuation in the
 momentum distribution. This may be done in a similar manner by employing
 the functional methods of Kadanoff and Baym\cite{Baym}.
 However, we shall instead adopt a simpler
 approach based upon some earlier work\cite{PREPRINT}. There we showed that
 for $ {\bf{k}} \neq {\bf{k}}^{'} $ that the number fluctuation has this
 rather simple looking form,
\begin{equation}
N({\bf{k}},{\bf{k}}^{'}) = \frac{1}{2}\langle n_{ {\bf{k}} } \rangle
n_{\beta}({\bf{k}}^{'}) +  \frac{1}{2}\langle n_{ {\bf{k}}^{'} } \rangle
n_{\beta}({\bf{k}}) - \langle n_{ {\bf{k}} } \rangle
\langle n_{ {\bf{k}}^{'} } \rangle
\end{equation}
where $ n_{\beta}({\bf{k}}) = 1/(exp(\beta(\epsilon_{ {\bf{k}} }-\mu_{0}))+1) $
and $ \mu_{0} $ is the chemical potential of noninteracting fermions
(at finite temperature). 
 This leads to a rather complicated set of self-consistent equations
 for the momentum distribution and its fluctuation. 
 These formulas have very appealing and illuminating
 forms and they also support our earlier conclusions\cite{Setlur}
 namely that Fermi liquid theory can break down
 in all three spatial dimensions for sufficiently strong values
 of the coupling and be intact for sufficiently weak couplings.
 This analysis also provides another (apart from our earlier preprint
 \onlinecite{preprint}) illustration of the importance of the concept
 of fluctuations in the momentum distribution. Lastly, it is worth remarking
 that we have not really answered the question posed in the title namely,
 is the random phase approximation(RPA) a controlled approximation ?
 It is our belief that the answer is no, for RPA cannot be thought of as the
 leading behaviour of an expansion in powers of an obvious small parameter. 
 The fact that the traditional RPA may be obtained as a weak coupling limit
 of Eq.(~\ref{GENRPA}) does not clarify the situation, for it may well
 be that even the generalised RPA is not a controlled approximation. 
 Indeed, we know for a fact that even the generalised RPA is not sufficient 
 in giving us a finite lifetime. Thus the title is somewhat rhetorical
 and it is hoped that critics of our earlier work will take note.
 This work was supported in part by ONR N00014-90-J-1267 and
 the Unversity of Illinois, Materials Research Laboratory under grant
 NSF/DMR-89-20539 and in part by the Dept. of Physics at
 University of Illinois at Urbana-Champaign. The authors may be contacted at
 the e-mail address setlur@mrlxpa.mrl.uiuc.edu.


\begin{thebibliography}{30}
\bibitem[9]{footnote} The pathological exception is for nonideal 
 distributions such as,
 $ \langle n_{ {\bf{k}} } \rangle = 0 $ for $ |{\bf{k}}| < 0.5\mbox{   }k_{f} $
 and $ \langle n_{ {\bf{k}} } \rangle = 1 $ for
 $ 0.5\mbox{   }k_{f} < |{\bf{k}}| < 1.5\mbox{   }k_{f} $
 and $ \langle n_{ {\bf{k}} } \rangle = 0 $ for
 $ |{\bf{k}}| > 1.5\mbox{   }k_{f} $ 
 this is very nonideal but has zero fluctuation, 
 but hopefully these are not physical.
\end{thebibliography}
\end{document}